\documentclass[a4paper]{article}

\usepackage{enumitem}

\usepackage[utf8]{inputenc}
\usepackage{graphicx}
\usepackage{float}
\usepackage{longtable}
\usepackage{biblatex}
\usepackage{xargs}
\usepackage{listings}
\makeatletter

\usepackage{jlcode}
\usepackage{hyperref}
\hypersetup{
    hidelinks,
}
\usepackage{cleveref}

\newcommand{\codeurl}{https://raw.githubusercontent.com/m3g/jlcode_example/master}
\newcommand{\codelinktext}{[Click here to download the code]}

\def\jlbasicfont{\ttfamily\footnotesize\selectfont}

\lstdefinestyle{code}{
    language=Julia, 
    showstringspaces=false,
    xleftmargin=5pt,
    framexleftmargin=10pt,
    linewidth = 1.08\linewidth,
    numbers=none,
    rulecolor        = \color[HTML]{000000},
    frame=false,
    keepspaces=true
}
\lstnewenvironment{code}[1][]{\lstset{
style=code, #1
}}{}
\newcommandx{\codefile}[3][1={}, 2={}]{\jlinputlisting[
    caption=#1,
    label=#2
]{#3}}
\newcommand{\codelink}[1]{\lstinputlisting[
    style=code
]{#1}
  \noindent\begin{center}
  \filename@parse{#1}
  \href{\codeurl/\filename@base.\filename@ext}
  {\textcolor{blue}{\codelinktext}}
  \end{center}
}
\newcommand{\linkonly}[1]{
  \protect\filename@parse{#1}
  \href{\codeurl/\filename@base.\filename@ext}
  {\textcolor{blue}{\tt[\protect\filename@base.\protect\filename@ext]}}
}
\makeatother
\newcommand{\inline}{\jlinl}

\newcommand{\Plots}{\texttt{Plots.jl}~}
\newcommand{\RecipesBase}{\texttt{RecipesBase.jl}~}

\begin{document}

\section*{(1) Overview}

\vspace{0.5cm}

\section*{Title}

\Plots -- a user extendable plotting API for the julia programming language

\section*{Paper Authors}

1. Christ, Simon; \\
2. Schwabeneder, Daniel; \\
3. Rackauckas, Christopher; \\
4. Borregaard, Michael Krabbe; \\
5. Breloff, Thomas

\section*{Paper Author Roles and Affiliations}
1. Leibniz Universit\"at Hannover\\
2. TU Wien \\
3. Massachusetts Institute of Technology \\
4. Center for Macroecology, Evolution and Climate, Globe Institute, University of Copenhagen \\
5. Headlands Technologies

\section*{Abstract}

There are plenty of excellent plotting libraries.
Each excels at a different use case: one is good for printed 2D publication figures, the other at interactive 3D graphics, a third has excellent \LaTeX{} integration or is good for creating dashboards on the web.

The aim of \Plots is to enable the user to use the same syntax to interact with many different plotting libraries, such that it is possible to change the library "backend" without needing to touch the code that creates the content – and without having to learn yet another application programming interface (API).

This is achieved by the separation of the plot specification from the implementation of the actual graphical backend.
These plot specifications may be extended by a "recipe" system, which allows package authors and users to define how to plot any new type (be it a statistical model, a map, a phylogenetic tree or the solution to a system of differential equations) and create new types of plots - without depending on the \Plots package.
This supports a modular ecosystem structure for plotting and yields a high reuse potential across the entire julia package ecosystem.
\Plots is publicly available at \url{https://github.com/JuliaPlots/Plots.jl}.

\section*{Keywords}

visualization; julia; plotting; julia-language; user-extendable

\section*{Introduction}

Julia\cite{bezansonJuliaFreshApproach2017a} is a programming language that achieves high performance and stellar modularity and composability by making use of multiple dispatch and just-in-time compilation.
This comes at the cost of increased latency as the language compiles new machine-code the first time any function is called on new types of arguments.
This is notoriously an issue for packages that call a large part of their codebase in the first call, such as plotting packages.
It even coined the term "time to first plot" as a phrase for julia's start-up latency.
Indeed, the julia language survey 2020\cite{shah2020JuliaUser} identified "it takes too long to generate the first plot" as the biggest problem faced by Julia users.

Package authors try to minimize loading time by reducing the number of dependencies, in particular those with long loading times themselves.
Thus, authors are faced with a challenge if they want to define new plotting functionality for their packages; e.g. if a package for differential equations wishes to make it possible for users to investigate different solutions visually.
Depending on a plotting package drastically increases startup times, as well as limiting users to that particular plotting package (which may conflict with other plotting packages used by the project).
As such, depending on plotting packages is rarely seen in the julia ecosystem.

\Plots has solved this problem, by introducing plotting "recipes", which allow package authors to only depend on a very lightweight package \inline{RecipesBase.jl} instead of depending on \Plots.
This package has no other effect than making specialized syntax available for the code author to define visualizations; but otherwise has no effect, until the package end user loads \Plots directly.
Thus, \Plots offers a unified and powerful API with a convenient way for package authors to support visualizations for multiple plotting packages, without increasing the loading time of their package – with the definition of a single recipe.
An example can be seen in listing~\ref{lst:measurements}.

\subsection*{Development}

\Plots was created by Tom Breloff between September 2015 and 2017, with the goal of creating a plotting API for the julia\cite{bezansonJuliaFreshApproach2017a} language, that was powerful, intuitive, concise, flexible, consistent, lightweight and smart.
In particular the recipe system helped the package gain large traction within the community, as the latency of loading large dependencies was generally recognized as one of the major factors limiting the uptake of Julia.

With time Tom moved on, and the development of \Plots was continued by Michael K. Borregaard and Daniel Schwabeneder.
The maintenance of the project is now a joint effort of the julia community.
The package has reached a very high uptake in the ecosystem.
In the Julia Language Survey of both 2019\cite{shahJuliaUserDeveloper} and 2020\cite{shah2020JuliaUser}, \Plots was identified as the julia community's favorite package across the entire ecosystem, with 47 percent of all julia users listing it among their favorite packages.

\subsection*{Usage}

\Plots is used for visualizations in scientific publications of different fields, such as numerics\cite{rackauckasDifferentialEquationsJlPerformant2017a,baggecarlsonMonteCarloMeasurementsJlPropagation2019,caldwellBATJlUpgrading2020,cufarRipsererJlFlexible2020,fairbrotherGaussianProcessesJlNonparametric2019,lindnerNetworkDynamicsJlComposing2021}, mathematics\cite{driscollComplexRegionsJlJulia2019}, biology\cite{angevaareInfectiousDiseaseTransmission2020,bonhamMicrobiomeJlBiobakeryUtils2021}, ecology\cite{dansereauSimpleSDMLayersJlGBIF2021} and geology\cite{constantinouGeophysicalFlowsJlSolvers2021,KellerSilicaCrust2020} as well as for teaching purposes\cite{boydIntroductionAppliedLinear2018,IntroductionComputationalThinking}.

Many packages in the julia ecosystem, as well as non-packaged code (e.g. for scientific projects and publications) contain \Plots recipes.
According to recent download statistics\cite{PackageDownloadStats} \Plots has between 500 and 2000 downloads per day, and >300 published packages in the general package registry of Julia currently have recipes for \Plots defined.

\subsection*{Comparison}
\Plots achieves its functionality by leveraging the multiple dispatch paradigm of julia, which allows the user to define multiple methods for the same function, with the compiler selecting the appropriate method based on the types of the input arguments.
Because of the close connection to Julia's multiple dispatch, it's approach to plotting is fairly unique.

In python, the library unified-plotting\cite{UnifiedPlottingUnifiedplotting} shares the aim of providing a unified API for multiple packages, in this case matplotlib\cite{Hunter:2007}, pyplot and javascript libraries including d3.js\cite{bostockD3JsDataDriven}.
However, unified-plotting is still in the beta phase and not widely used.

The authors are not aware of other package ecosystems that have a recipe system akin to that of \Plots{}, though a recipe system inspired by that of \Plots{} is presently being implemented for the julia library \inline{Makie.jl}\cite{danischMakieJlFlexible2021}.

\section*{Implementation and architecture}

\subsection*{One-function API\footnotemark}

\footnotetext{Technically the API consists of more than one function, but the vast majority is \texttt{plot/plot!} and aliases thereof.}
A central design goal of \Plots is that the user should rarely have to consult the documentation while plotting.
This is achieved by having a tightly unified syntax.
\Plots's main interface is simply the \texttt{plot} function, which creates a new plot object.
Additionally there is the \texttt{plot!} function to modify an existing plot object, e.g. by changing axes limits or adding new elements.
Any type of predefined plot (e.g. a histogram, bar plot, scatter plot, a heatmap, an image, a geographical map etc.), may be created by a call to \texttt{plot} - the exact type is defined by the keyword argument \inline{seriestype} and the input arguments (type and number).
New seriestypes can be created with recipes (see below).

For convenience, \Plots also exports "shorthand" functions named after the seriestypes (see examples in listing~\ref{lst:shorthands}).
\pagebreak
\begin{code}[caption=Examples of shorthands. Full list available at \url{https://docs.juliaplots/stable/api/\#Plot-specification}., label=lst:shorthands]
boxplot(args...; kwargs...) = plot(args...; seriestype = :boxplot, kwargs...)
scatter(args...; kwargs...) = plot(args...; seriestype = :scatter, kwargs...)
\end{code}

\begin{figure}
    \centering
    \includegraphics[width=\linewidth]{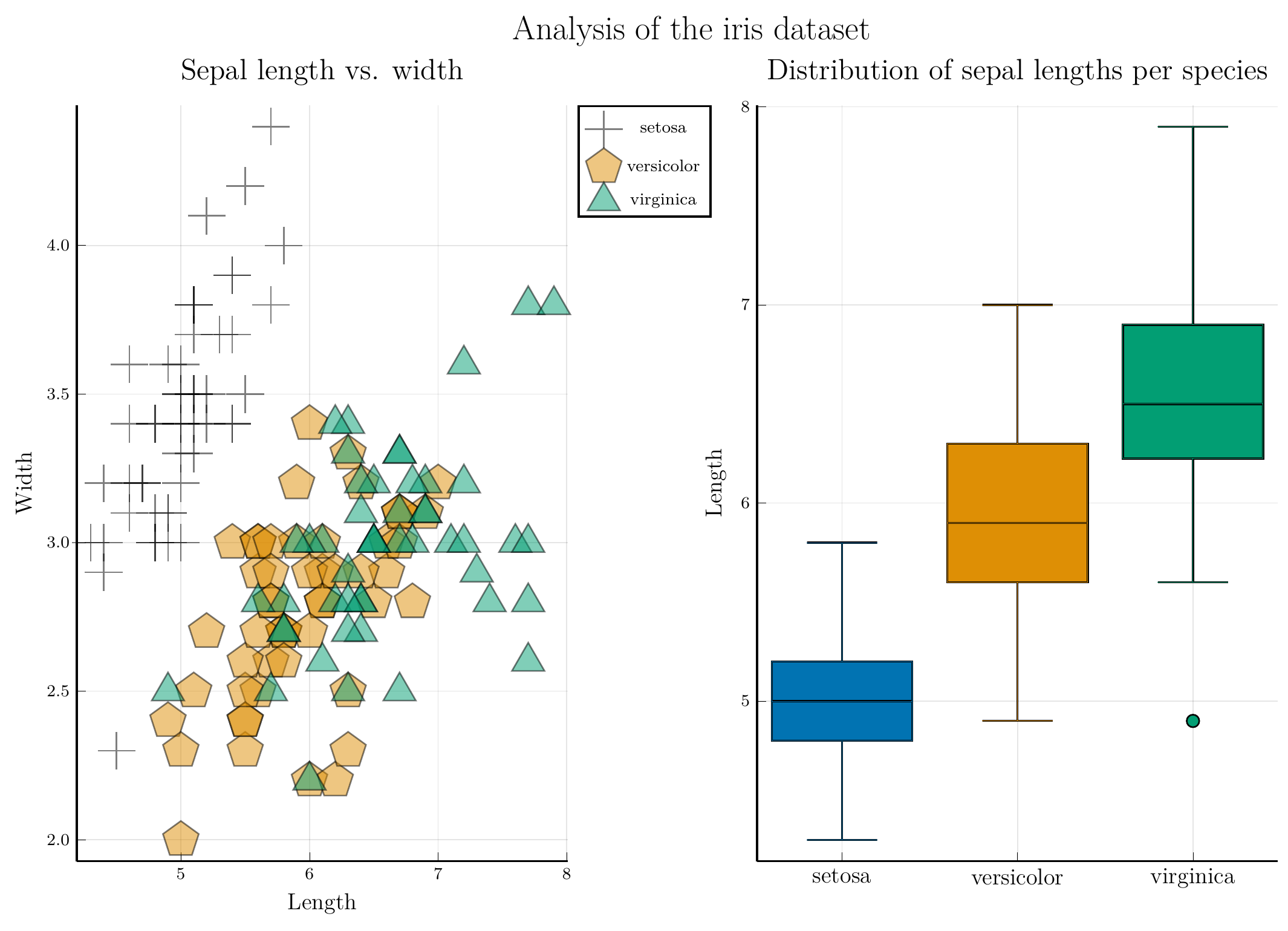}
    \caption{Example plot of the iris dataset\cite{UCIMachineLearning} to illustrate the use of different attribute types (cf. listing~\ref{lst:attributes}).}
    \label{fig:attributes}
\end{figure}

All aspects of the plot are controlled by a set of plot \emph{attributes}, that are controlled by keyword arguments\cite{OverviewPlots}.
\Plots distinguishes four hierarchical levels of attributes: plot attributes, subplot attributes, axis attributes and series attributes (cf. \cref{fig:attributes}).

\jlinputlisting[caption=Code corresponding to \cref{fig:attributes}., label=lst:attributes]{./fig/attributes.jl}

A \emph{series} in a \Plots context is an individual plot element, such as a continuous line or a set of scatter points.
A plot may contain multiple series, e.g. when adding a trend line to a scatter plot.
Multiple series may be added in the same \inline{plot} call by concatenating the data as columns in a row matrix (see below).

Input arguments can have many different forms like:
\begin{code}
plot()                                       # empty Plot with axes
plot(4)                                      # initialize a Plot with 4 empty series
plot(rand(10))                               # 1 series... x = 1:10
plot(rand(10,5), rand(10))                   # 5 series... y is the same for all
plot(sin, rand(10))                          # y = sin.(x)
plot([sin,cos], 0, pi)                       # sin and cos lines on the range [0, pi]
                                             # using an automatic adaptive grid
plot(1:10, Any[rand(10), sin])               # 2 series, y is rand(10) and sin.(x)
plot( plot(rand(10)), plot(rand(10)) )       # a layout with two equally sized
                                             # subplots
@df dataset("Ecdat", "Airline") plot(:Cost)  # the :Cost column from a DataFrame
                                             # @df is currently in StatsPlots.jl
\end{code}
Calling the \inline{plot} function returns a \inline{Plot} object.
The \texttt{Plot} object is essentially a big nested dictionary holding the plot attributes for the layout, subplots, series, segments, etc. and their values.
The plot object is automatically rendered in the surrounding context when returned to an interactive session, or can be displayed explicitly by calling the \texttt{display} function on the object.
This delayed rendering means that \texttt{plot} calls can be combined without unnecessary intermediate rendering.

\subsection*{Pipeline}

\begin{figure}[H]
    \centering
    \includegraphics[width=\textwidth]{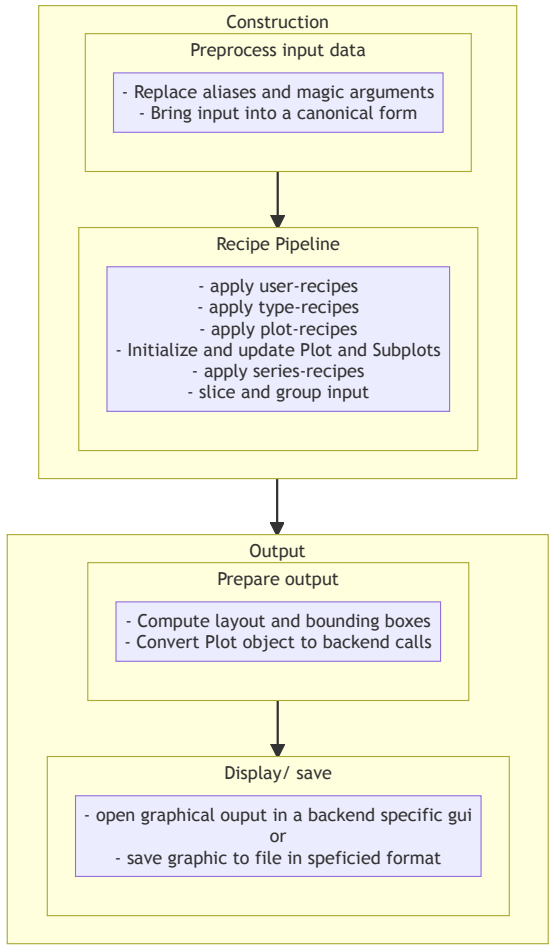}
    \caption{
        Plotting pipeline in \Plots{}.
        The separation of construction and output production enables the flexible use of different backends in the same session and helps to avoid unnecessary intermediate calculation.
        Created using \texttt{mermaid}\cite{Mermaid}.
    }
    \label{fig:pipeline}
\end{figure}

The plotting pipeline mainly has two stages (cf. \cref{fig:pipeline}): construction of the plot using \inline{plot/plot!} calls and creating the output via \inline{savefig/display/gui} calls.
These calls are often called implicitly in environments like the julia REPL, notebooks or IDEs.

The very first step upon construction is to convert all inputs to form the list of plot attributes that constitute the plot specification.
As shown in listing~\ref{lst:aliases} \Plots is very flexible about possible input values.
The conversion step involves defining values for all attributes based on the values input as keyword arguments.
This includes replacing "aliases" of attributes (which are multiple alternatively spelled keywords, such as `c` or `color`, encoding the same attribute), handling of \inline{missing} and \inline{nothing} values in the input data and attribute values, and determining the final values based on the set of defaults.
The default values are organized in a hierarchical framework, based on the values of other attributes; e.g. \texttt{linecolor}, \texttt{fillcolor} and \texttt{markercolor} will default to \texttt{seriescolor} under most seriestypes.
But, for instance, under the \texttt{bar} seriestype, \texttt{linecolor} will default to \texttt{:black}, giving bars with a black border.
This allows the specification of useful plots with a minimum of specification, in contrast to the paradigm of e.g. matplotlib, where every aspect of the plot is usually defined manually by the user.

\begin{code}[caption=Examples of input preprocessing steps in \Plots. All these calls are equivalent., label=lst:aliases]
plot(2:4, c = :steelblue)                # c is the shortest alias for seriescolor
plot([2,3,4], color = 1)                 # :steelblue is the first color
                                         # of the default palette
plot(1:3, [2,3,4], colour = :auto)       # the recipe for a single input
                                         # will use 1:3 as x-values
plot(1:3, [2,3,4], seriescolors = 1)     # you can use singular
                                         # or plural version of attributes
plot([1,2,3], [2,3,4], seriescolor = RGBA{Float64}(0.275,0.51,0.706,1.0))
# this is the fully expanded call
\end{code}

Afterwards recipes are applied recursively and the \inline{Plot} and \inline{Subplot} objects are initialized.
Recipes will be explained in detail in the next section.

When an output is to be produced the layout will be computed and the backend-specific code will be executed to produce the result.

\subsection*{Recipes}
\label{sec:recipes}

As mentioned in the introduction, recipes are the key mechanism in the \Plots \linebreak pipeline to allow composable definitions of visualisations across julia packages.
The composable definitions may be applied recursively, which is a major advancement for improving ecosystem support by giving a combinatoric reduction in the amount of code required for downstream libraries to add native plotting support for their types.

\Plots distinguishes four types of recipes: user recipes, type recipes, plot recipes and series recipes \cite{HowRecipesActually}.
User recipes (which define how to plot objects of a certain type) and series recipes (which define a new seriestype) are by far the most commonly used.
All of them can be constructed with the \inline{@recipe} macro which acts on a function definition.
The type of the recipe is then determined by the signature of that function, utilizing the multiple dispatch capabilities of the julia programming language.

\begin{code}[caption=Recipe signatures, label=lst:recipe_signature]
using RecipesBase
struct CustomStruct end
@recipe function f(arg::CustomStruct; custom_kw = 1)            # user recipe
end
@recipe function f(::Type{CustomStruct}, val::CustomStruct)     # type recipe
end
@recipe function f(::Type{Val{:recipename}}, plt::AbstractPlot) # plot recipe
end
@recipe function f(::Type{Val{:recipename}}, x, y, z)           # series recipe
end
\end{code}

It is enough to depend on the \inline{RecipesBase.jl} package, a small and lightweight dependency to define a recipe.

The major question with recipes is how this is an improvement over previous designs.
For example, in most plotting libraries such as \texttt{matplotlib}\cite{Hunter:2007}, a downstream ODE solver library can add a new function \inline{plotsolution} that will plot an ODE solution.
However, the difference, and the major technological advance of the \Plots recipe system, is that the application of recipes is recursive and extendable via multiple dispatch.
This solves a combinatoric problem for downstream support: it is possible to combine and chain recipes to support plotting on new combinations of input types without ever defining a recipe for that specific combination.

\begin{figure}[H]
    \centering
    \includegraphics[width=\textwidth]{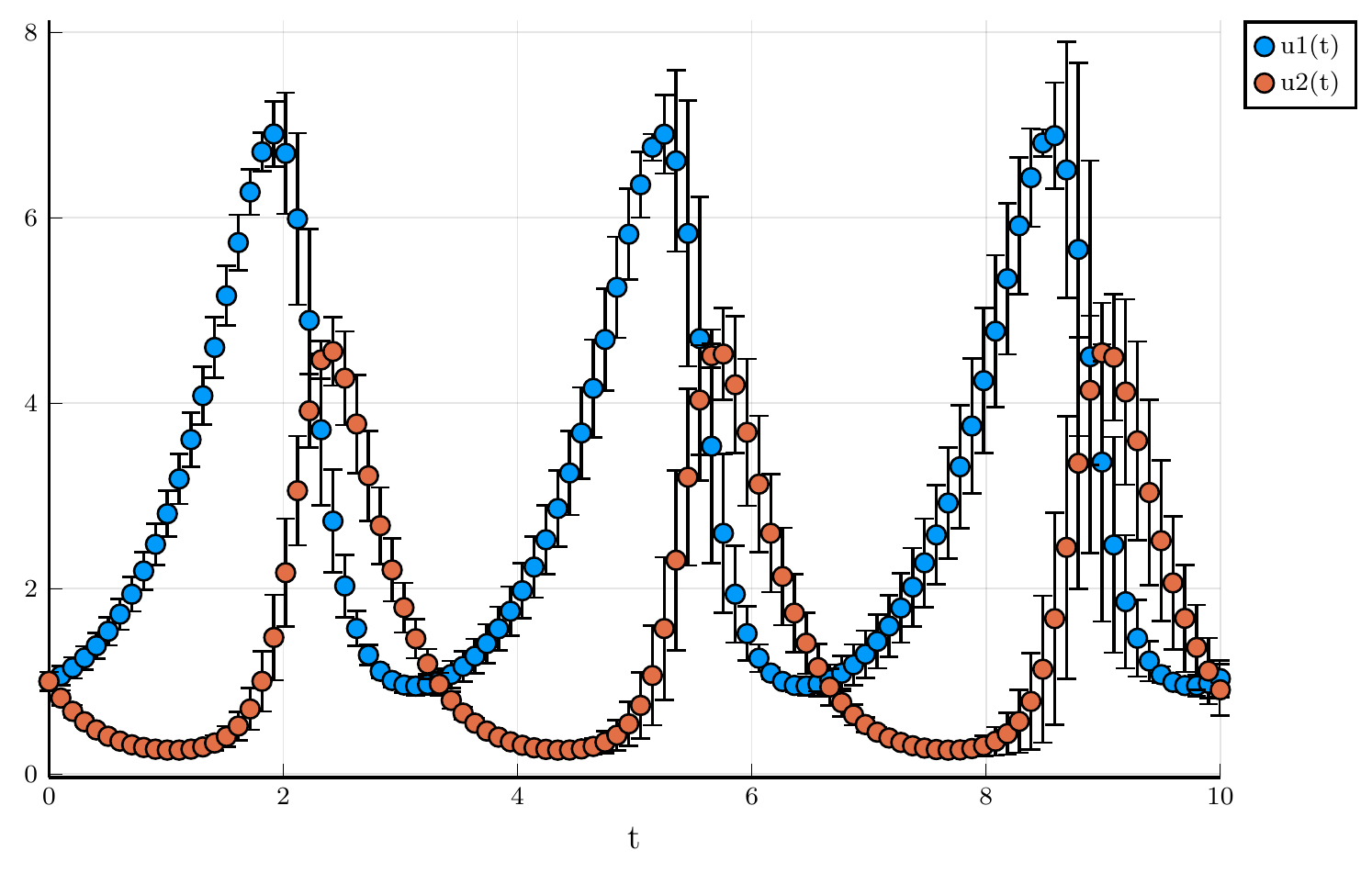}
    \caption{
        Showcase of composing recipes.
        Plotting a \texttt{ODESolution} object from \texttt{DifferentialEquations.jl} containing \texttt{Measurement}s from \texttt{Measurements.jl} will apply the recipe of \texttt{DifferentialEquations.jl} which will return vectors of \texttt{Measurement}s, which will apply the recipe from \texttt{Measurements.jl}; yielding the solutions of the Lotka-Volterra system\cite{alfredj.lotkaElementsPhysicalBiology1925} with correct error bounds without the user having to change the callsite.
        Neither of these packages has code in their recipes for handling types of the other package.
        Full code available in listing~\ref{lst:recipe}.
    }
    \label{fig:DiffEqM}
\end{figure}

To illustrate this, consider the example of combining the recipes defined by the julia packages\linebreak \inline{DifferentialEquations.jl}\cite{rackauckasSciMLDifferentialEquationsJl2022} and \inline{Measurements.jl}\cite{giordanoUncertaintyPropagationFunctionally2016} (cf.~\cref{fig:DiffEqM} and listing~\ref{lst:recipe}).
In this example, a user solves a differential equation with uncertain initial conditions specified by \inline{Measurements.Measurement} objects.
The uncertainty encoded in the Measurement objects are automatically propagated through the ODE solver, as multiple methods for this type have been defined for the arithmetic functions.
The resulting ODE solution \inline{sol} is then already specified in terms of such \linebreak\inline{Measurements.Measurement}s.
When running the plot command \inline{plot(sol)}, the recipe for ODE solvers will transform the \inline{ODESolution} object into an array of arrays, each representing a time series to plot (using techniques like dense output to produce a continuous looking solution).
This array of arrays contains number types matching the state of the solution, in this case \inline{Measurements.Measurement}s.
Successive applications of the user recipe defined in \inline{Measurements.jl} then take each state value and assign the \inline{uncertainty} part of the state to the \inline{yerror} attribute and pass the \inline{value} part of the state to the next recipe.
When used with the initial seriestype \texttt{:scatter} this results in a scatter plot with proper error bars as seen in \cref{fig:DiffEqM}.

Therefore, while the two packages were not developed to work together, multiple dispatch allows to  efficiently solve problems containing combinations of these packages, and the \Plots recipe system allows the combined visualization to work automatically.

The recipe of \inline{Measurements.jl} is an example of a particularly short recipe.
A \inline{Measurements.Measurement} is represented as a type with two fields: \inline{value} and\linebreak \inline{uncertainty}.
It can be conveniently constructed with the Unicode infix operator $\pm$.
Thus the object  $a \pm b$ has $a$ as the \inline{value} and $b$ as the \inline{uncertainty}.
An array of measurement values can be converted into an array of floating point values to plot, along with having the uncertainties as error bars, via the following recipe:

\begin{code}[caption={\texttt{Measurements.jl} recipe}, label={lst:measurements}]
@recipe function f(x::AbstractArray, y::AbstractArray{<:Measurement})
    yerror := uncertainty.(y)       # := is special syntax in the @recipe block which
                                    # sets an attribute overriding
                                    # any present value. The alternative syntax
                                    # is --> to give call-site values precedence.
    x, value.(y)
end
\end{code}

\subsection*{Structure and interfaces}

\begin{figure}[H]
    \centering
    \includegraphics[width=\textwidth]{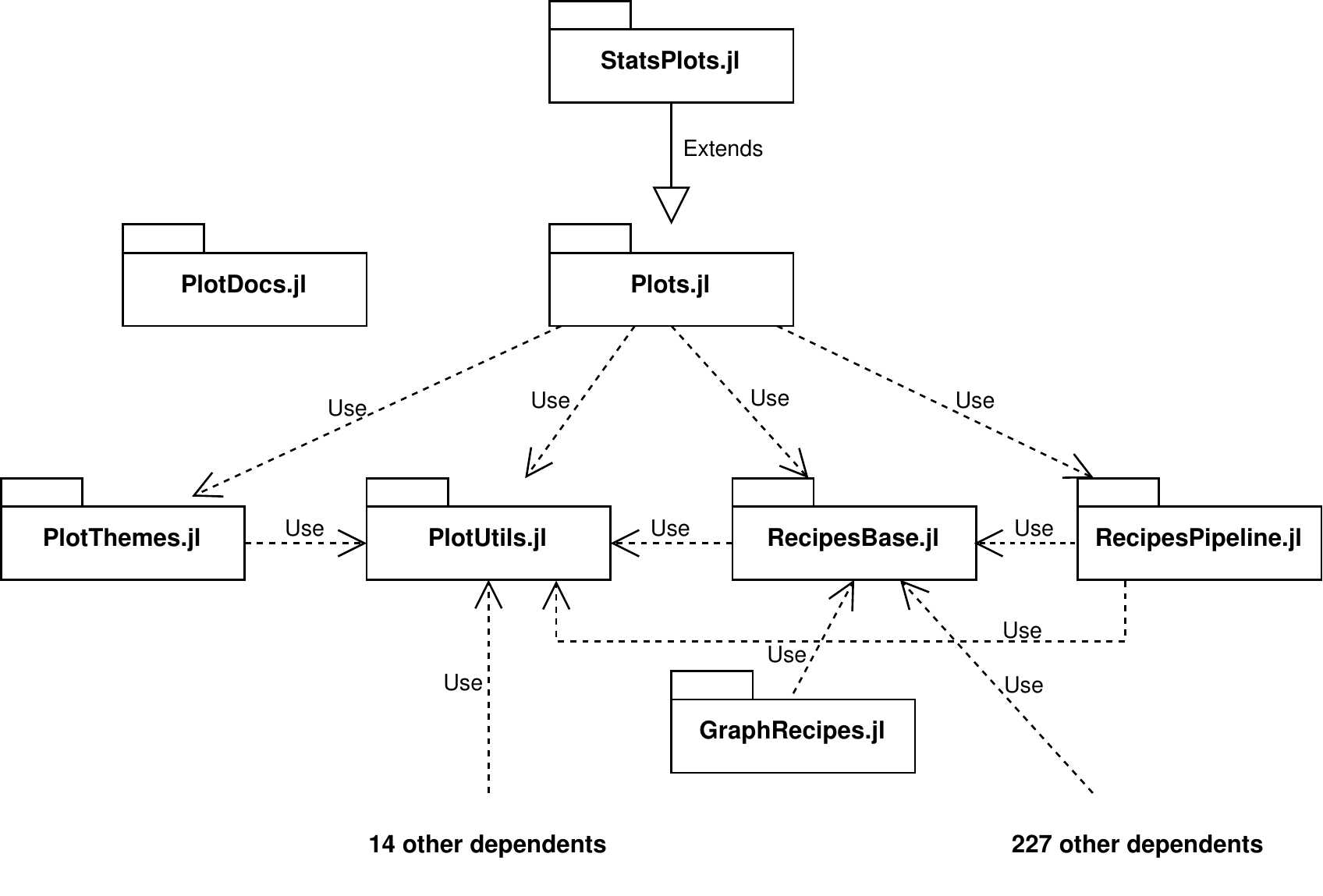}
    \caption{
        Overview of the \Plots ecosystem and its interfaces with other julia packages.
        The numbers of dependents are taken from juliahub\cite{PlotsJuliaHub}.
    }
    \label{fig:ecosystem}
\end{figure}

The code for \Plots is not located in one repository, but split into a few packages, to enhance reuse of more general parts of the code by other packages (cf.~\cref{fig:ecosystem}).
In the following the different packages and their use cases will be described.

\paragraph{\Plots:} The main user facing package.
Defines all default values and holds the code for layouting, conversion of input arguments, output generation, all backend code and the default recipes.
This is the repository with the highest rate of change.

\paragraph{\texttt{StatsPlots.jl}:} A drop-in replacement for \Plots, meaning it loads and reexports all of \Plots and adds recipes that are specially targeted at visualisation of statistical data (aiming to be integrated with Julia's statistical package ecosystem under the JuliaStats organisation).
Therefore it has more dependencies than \Plots which increases the loading time and since not all users need this functionality it is separated in its own repository.

\paragraph{\texttt{PlotUtils.jl}:} Provides general utility routines, such as handling colors, optimizing ticks or function sampling. This package is also used by e.g. the newer plotting package \inline{Makie.jl}.

\paragraph{\texttt{RecipesBase.jl}:} A package with zero 3rd-party dependencies, that can be used by other packages to define recipes for their own types without needing to depend on \Plots.

\paragraph{\texttt{RecipesPipeline.jl}:} Another lightweight package that defines an API such that other plotting packages can consume recipes from \RecipesBase without needing to become a backend of \Plots.

\paragraph{\texttt{GraphRecipes.jl}:} A package that provides recipes for visualisation of graphs in the sense of graph theory.
These are also split out because they have some heavy dependencies.

\paragraph{\texttt{PlotThemes.jl}:} Provides different themes for \Plots.

\paragraph{\texttt{PlotDocs.jl}:} Hosts the documentation of \Plots.

\subsection*{Backends}

\Plots currently supports seven plotting frameworks as backends.
Typically these plotting frameworks themselves have different graphic libraries as backends to support different output types.
The backends of \Plots differ in their area of expertise and have different trade-offs.

\paragraph{GR:} The default backend.
Uses the GR framework\cite{heinenGRFrameworkGR}.
It is among the fastest backends with a good coverage of functionality.

\paragraph{Plotly/PlotlyJS:} Is the backend with the most interactivity and best web support using the \texttt{plotly} javascript library\cite{PlotlyJavaScriptGraphing}.
One usecase is to create interactive plots in documentation\cite{PlottingSpectralDistances} or notebooks.
The \inline{Plotly} backend is a version with minimal dependencies, which doesn't require the user to load any other julia package and displays its graphics in the browser, while \inline{PlotlyJS} requires the user to load \inline{PlotlyJS.jl}, but offers display of plots in a standalone window.

\paragraph{PyPlot:} \texttt{PyPlot.jl} is the julia wrapper of \texttt{matplotlib}\cite{Hunter:2007} and covers a lot of functionality at moderate speed.

\paragraph{PGFPlotsX:} Uses the \texttt{pgfplots} \LaTeX package\cite{PGFPlotsLaTeXPackage} and is thus the slowest of the backends, but integrates very good with \LaTeX-documents.

\paragraph{InspectDR:} Fast backend with GUI and some interactivity that does good for 2D and handles large datasets and high refresh rates\cite{ma-laforgeInspectDRJlFast2022}.

\paragraph{UnicodePlots:} A backend that allows plotting in the terminal with unicode characters and can be used in a terminal (also on headless machines)\cite{UnicodePlots2022}.
Therefore it lacks a lot of functionality compared to the other backends.

\paragraph{HDF5:}
A backend that can be used to save the \texttt{Plot} object along the data in a hdf5-file using \texttt{HDF5.jl}\cite{HomeHDF5Jl}, such that it can be recovered with any backend.
Potentially allows interfacing with \Plots from other programming languages.

\vspace{1em}
Furthermore there are 6 deprecated backends that were used in the earlier stages of \Plots, but which are no longer maintained and the \inline{Gaston.jl} backend which is in an early experimental stage.
\inline{Gaston.jl} is a julia interface for \texttt{gnuplot}\cite{GnuplotHomepage}.
This shows that \Plots can be sustained even if a maintainer of backend code leaves.
Either the backend will be maintained by the community or it will be replaced by another backend.

\section*{Quality control}

\Plots runs its unit tests of all backends as well as visual regression tests of the default backend against the latest version of macOS, Ubuntu and Windows using the current stable version of julia, the long term support version and the nightly version on every pull request and pushes to the default branch.
Furthermore benchmarks are run to detect performance regressions.
Lastly, building the documentation creates a suite of example plots for every backend, which would also detect certain errors.

\section*{(2) Availability}
\vspace{0.5cm}
\section*{Operating system}

\Plots is tested on Windows, Linux and macOS.

\section*{Programming language}

julia 1.5

\section*{Additional system requirements}

\section*{Dependencies}

\Plots has the following direct dependencies:
\begin{description}
    \item[Contour.jl] v0.5
    \item[FFMPEG.jl] v0.2 - v0.4
    \item[FixedPointNumbers] v0.6 - v0.8
    \item[GR.jl] v0.46 - v0.55, v0.57
    \item[GeometryBasics.jl] v0.2, v0.3.1 - v0.3
    \item[JSON.jl] v0.21, v1
    \item[Latexify.jl] v0.14 - v0.15
    \item[Measures.jl] v0.3
    \item[NaNMath.jl] v0.3
    \item[PlotThemes.jl] v2
    \item[PlotUtils.jl] v1
    \item[RecipesBase.jl] v1
    \item[RecipesPipeline.jl] v0.3
    \item[Reexport.jl] v0.2, v1
    \item[Requires.jl] v1
    \item[Scratch.jl] v1
    \item[Showoff.jl] v0.3.1 - v0.3, v1
    \item[StatsBase.jl] v0.32 - v0.33
\end{description}

In addition it has 125 indirect dependencies all of which can be seen at \cite{PlotsJuliaHub}.

\section*{List of contributors}

\begin{figure}[H]
    \centering
    \includegraphics[width=\textwidth]{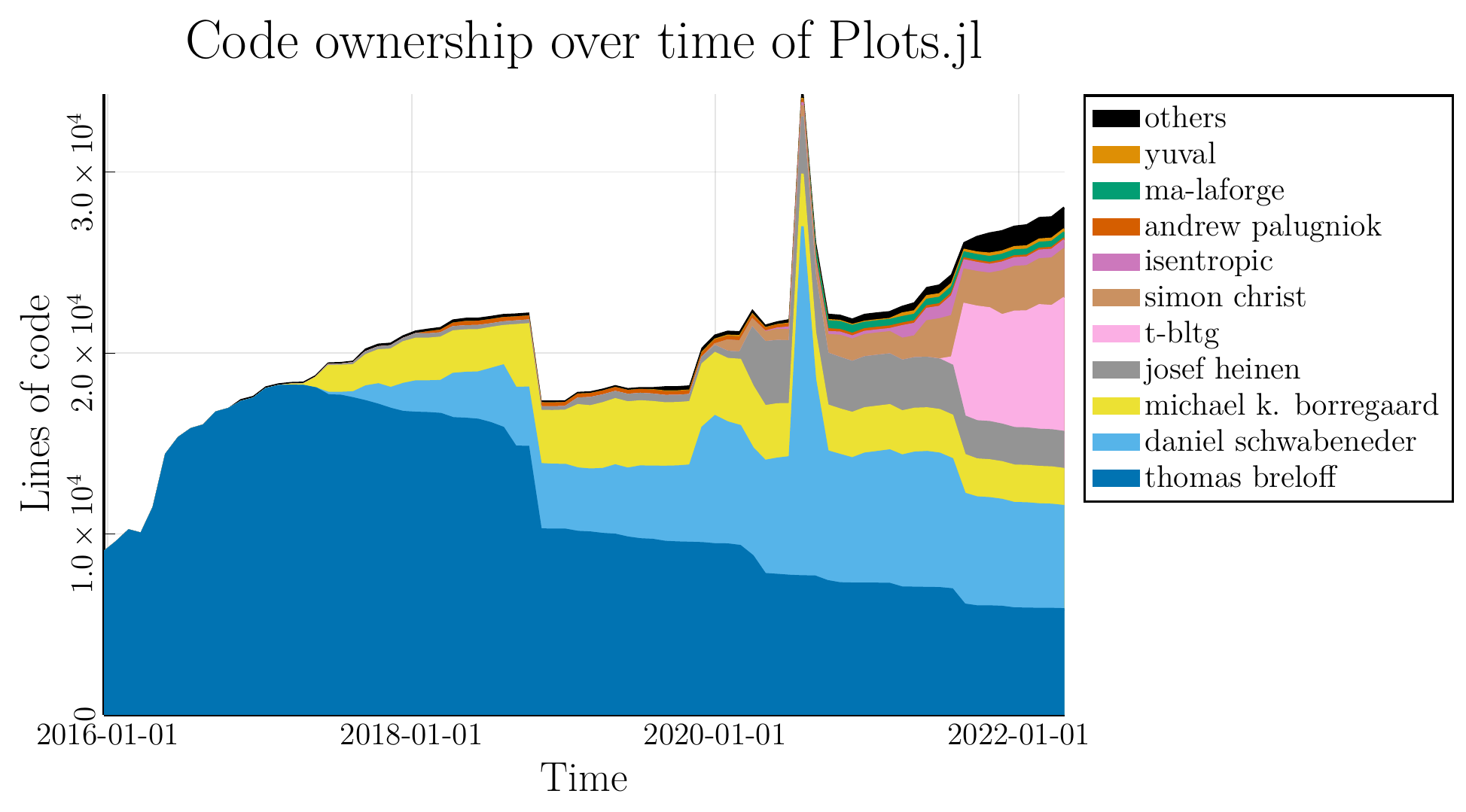}
    \caption{
        Lines of code alive of the top ten contributors of the \Plots repository over time.
        Data created with \texttt{hercules}\cite{SrcdHercules2021}.
        }
    \label{fig:burndown_people}
\end{figure}

\begin{figure}[H]
    \centering
    \includegraphics[width=\textwidth]{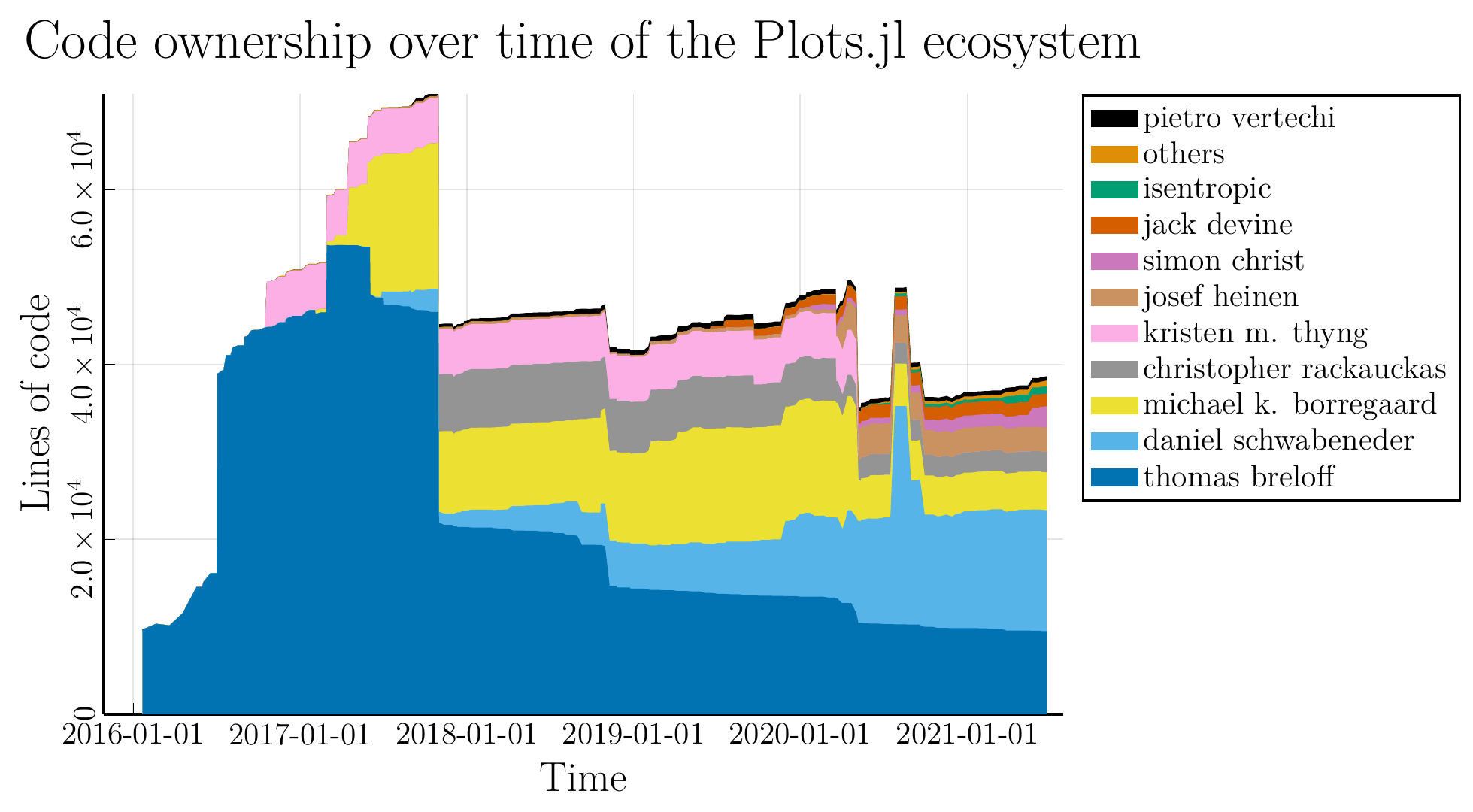}
    \caption{
        Lines of code alive of the top ten contributors of the \Plots ecosystem (\cref{fig:ecosystem}) over time.
        Data created with \texttt{hercules}\cite{SrcdHercules2021}.
        }
    \label{fig:burndown_eco}
\end{figure}

\begin{longtable}{p{4cm}p{4cm}ll}
    \caption{Contributors sorted by number of commits.}\\
  \hline\hline
  \textbf{name} & \textbf{affiliation} & \textbf{role} & \textbf{orcid} \\\hline
  \endfirsthead
  \hline\hline
  \textbf{name} & \textbf{affiliation} & \textbf{role} & \textbf{orcid} \\\hline
  \endhead
  \hline\hline
  \endfoot
  \endlastfoot
  Tom Breloff & Headlands Technologies & Creator & missing \\
  Daniel Schwabeneder & TU Wien & ProjectLeader & 0000-0002-0412-0777 \\
  Michael Krabbe Borregaard & GLOBE Institute, University of Copenhagen & ProjectLeader & 0000-0002-8146-8435 \\
  Simon Christ & Leibniz Universität Hannover & ProjectLeader & 0000-0002-5866-1472 \\
  Josef Heinen & Forschungszentrum Jülich & ProjectMember & 0000-0001-6509-1925 \\
  Yuval & missing & Other & missing \\
  Andrew Palugniok & missing & ProjectMember & missing \\
  Simon Danisch & @beacon-biosignals & Other & missing \\
  Pietro Vertechi & Veos Digital (https://veos.digital/) & ProjectMember & missing \\
  Zhanibek Omarov & Korea Advanced Inst. of Science and Technology (KAIST) & ProjectMember & 0000-0002-8783-8791 \\
  Thatcher Chamberlin & missing & Other & missing \\
  @ma-laforge & missing & ProjectMember & missing \\
  Christopher Rackauckas & Massachusetts Institute of Technology & Other & 0000-0001-5850-0663 \\
  Oliver Schulz & Max Planck Institute for Physics & Other & missing \\
  Sebastian Pfitzner & @JuliaComputing & Other & missing \\
  Takafumi Arakaki & missing & Other & missing \\
  Amin Yahyaabadi & University of Manitoba & Other & missing \\
  Jack Devine & missing & Other & missing \\
  Sebastian Pech & missing & Other & missing \\
  Patrick Kofod Mogensen & @JuliaComputing & Other & 0000-0002-4910-1932 \\
  Samuel S. Watson & missing & Other & missing \\
  Naoki Saito & UC Davis & Other & 0000-0001-5234-4719 \\
  Benoit Pasquier & University of Southern California (USC) & Other & 0000-0002-3838-5976 \\
  Ronny Bergmann & NTNU Trondheim & Other & 0000-0001-8342-7218 \\
  Andy Nowacki & University of Leeds & Other & 0000-0001-7669-7383 \\
  Ian Butterworth & missing & Other & missing \\
  David Gustavsson & Lund University & Other & missing \\
  Anshul Singhvi & Columbia University & Other & 0000-0001-6055-1291 \\
  david-macmahon & missing & Other & missing \\
  Fredrik Ekre & missing & Other & missing \\
  Maaz Bin Tahir Saeed & missing & Other & missing \\
  Kristoffer Carlsson & missing & Other & missing \\
  Will Kearney & missing & Other & missing \\
  Niklas Korsbo & missing & Other & missing \\
  Miles Lucas & missing & Other & missing \\
  @Godisemo & missing & Other & missing \\
  Florian Oswald & missing & Other & missing \\
  Diego Javier Zea & missing & Other & missing \\
  @WillRam & missing & Other & missing \\
  Fedor Bezrukov & missing & Other & missing \\
  Spencer Lyon & missing & Other & missing \\
  Darwin Darakananda & missing & Other & missing \\
  Lukas Hauertmann & missing & Other & missing \\
  Huckleberry Febbo & missing & Other & missing \\
  @H-M-H & missing & Other & missing \\
  Josh Day & missing & Other & missing \\
  @wfgra & missing & Other & missing \\
  Sheehan Olver & missing & Other & missing \\
  Jerry Ling & missing & Other & missing \\
  Jks Liu & missing & Other & missing \\
  Seth Axen & missing & Other & missing \\
  @o01eg & missing & Other & missing \\
  Sebastian Micluța-Câmpeanu & missing & Other & missing \\
  Tim Holy & missing & Other & missing \\
  Tony Kelman & missing & Other & missing \\
  Antoine Levitt & missing & Other & missing \\
  Iblis Lin & missing & Other & missing \\
  Harry Scholes & missing & Other & missing \\
  @djsegal & missing & Other & missing \\
  Goran Nakerst & missing & Other & missing \\
  Felix Hagemann & missing & Other & missing \\
  Matthieu Gomez & missing & Other & missing \\
  @biggsbiggsby & missing & Other & missing \\
  Jonathan Anderson & missing & Other & missing \\
  Michael Kraus & missing & Other & missing \\
  Carlo Lucibello & missing & Other & missing \\
  Robin Deits & missing & Other & missing \\
  Misha Mkhasenko & missing & Other & missing \\
  Benoît Legat & missing & Other & missing \\
  Steven G. Johnson & missing & Other & missing \\
  John Verzani & missing & Other & missing \\
  Mattias Fält & missing & Other & missing \\
  Rashika Karki & missing & Other & missing \\
  Morten Piibeleht & missing & Other & missing \\
  Filippo Vicentini & missing & Other & missing \\
  David Anthoff & missing & Other & missing \\
  Leon Wabeke & missing & Other & missing \\
  Yusuke Kominami & missing & Other & missing \\
  Oscar Dowson & missing & Other & missing \\
  Max G & missing & Other & missing \\
  Fabian Greimel & missing & Other & missing \\
  Jérémy & missing & Other & missing \\
  Pearl Li & missing & Other & missing \\
  David P. Sanders & missing & Other & missing \\
  Asbjørn Nilsen Riseth & missing & Other & missing \\
  Jan Weidner & missing & Other & missing \\
  @jakkor2 & missing & Other & missing \\
  Pablo Zubieta & missing & Other & missing \\
  Hamza Yusuf Çakır & missing & Other & missing \\
  John Rinehart & missing & Other & missing \\
  Martin Biel & missing & Other & missing \\
  Moritz Schauer & missing & Other & missing \\
  Mosè Giodano & missing & Other & missing \\
  @olegshtch & missing & Other & missing \\
  Leon Shen & missing & Other & missing \\
  Jeff Fessler & missing & Other & missing \\
  @hustf & missing & Other & missing \\
  Asim H Dar & missing & Other & missing \\
  @8uurg & missing & Other & missing \\
  Abel Siqueira & missing & Other & missing \\
  Adrian Dawid & missing & Other & missing \\
  Alberto Lusiani & missing & Other & missing \\
  Balázs Mezei & missing & Other & missing \\
  Ben Ide & missing & Other & missing \\
  Benjamin Lungwitz & missing & Other & missing \\
  Bernd Riederer & University of Graz & Other & 0000-0001-8390-0087 \\
  Christina Lee & missing & Other & missing \\
  Christof Stocker & missing & Other & missing \\
  Christoph Finkensiep & missing & Other & missing \\
  @Cornelius-G & missing & Other & missing \\
  Daniel Høegh & missing & Other & missing \\
  Denny Biasiolli & missing & Other & missing \\
  Dieter Castel & missing & Other & missing \\
  Elliot Saba & missing & Other & missing \\
  Fengyang Wang & missing & Other & missing \\
  Fons van der Plas & missing & Other & missing \\
  Fredrik Bagge Carlson & missing & Other & missing \\
  Graham Smith & missing & Other & missing \\
  Hayato Ikoma & missing & Other & missing \\
  Hessam Mehr & missing & Other & missing \\
  @InfiniteChai & missing & Other & missing \\
  Jack Dunn & missing & Other & missing \\
  Jeff Bezanson & missing & Other & missing \\
  Jeff Eldredge & missing & Other & missing \\
  Jinay Jain & missing & Other & missing \\
  Johan Blåbäck & missing & Other & missing \\
  @jmert & missing & Other & missing \\
  Lakshya Khatri & missing & Other & missing \\
  Lia Siegelmann & missing & Other & missing \\
  @marekkukan-tw & missing & Other & missing \\
  Mauro Werder & ETH Zurich & Other & 0000-0003-0137-9377 \\
  Maxim Grechkin & missing & Other & missing \\
  Michael Cawte & missing & Other & missing \\
  @milesfrain & missing & Other & missing \\
  Nicholas Bauer & missing & Other & missing \\
  Nicolau Leal Werneck & missing & Other & missing \\
  @nilshg & missing & Other & missing \\
  Oliver Evans & missing & Other & missing \\
  Peter Gagarinov & missing & Other & missing \\
  Páll Haraldsson & missing & Other & missing \\
  Rik Huijzer & missing & Other & missing \\
  Romain Franconville & missing & Other & missing \\
  Ronan Pigott & missing & Other & missing \\
  Roshan Shariff & missing & Other & missing \\
  Scott Thomas & missing & Other & missing \\
  Sebastian Rollén & missing & Other & missing \\
  Seth Bromberger & missing & Other & missing \\
  Siva Swaminathan & missing & Other & missing \\
  Tim DuBois & missing & Other & missing \\
  Travis DePrato & missing & Other & missing \\
  Will Thompson & missing & Other & missing \\
  Yakir Luc Gagnon & missing & Other & missing \\
  Benjamin Chislett & missing & Other & missing \\
  @hhaensel & missing & Other & missing \\
  @improbable22 & missing & Other & missing \\
  Johannes Fleck & missing & Other & missing \\
  Peter Czaban & missing & Other & missing \\
  @innerlee & missing & Other & missing \\
  Mats Cronqvist & missing & Other & missing \\
  Shi Pengcheng & missing & Other & missing \\
  @wg030 & missing & Other & missing \\
  Will Tebbutt & University of Cambridge & Other & missing \\
  @t-bltg & missing & Other & missing \\
  Fred Callaway & missing & Other & missing \\
  Jan Thorben Schneider & missing & Other & missing \\
  Lee Phillips & Alogus Research Corporation & Other & 0000-0003-4102-2460 \\
  Tom Gillam & missing & Other & missing \\\hline\hline
\end{longtable}

The code for creating tables and figures is publicly available at \url{https://gitlab.uni-hannover.de/comp-bio/manuscripts/plots-paper}.

\section*{Software location:}

{\bf Code repository} Github

\begin{itemize}[noitemsep,topsep=0pt]
	\item[Name:] JuliaPlots/Plots.jl
	\item[Persistent identifier:] \url{https://doi.org/10.5281/zenodo.4725318}
	\item[Licence:] MIT
        	\item[Version published:] 1.13.2
	\item[Date published:] 28/04/2021
\end{itemize}

The first version of \Plots was published on github at 11/09/2015.

\section*{Language}

julia

\section*{(3) Reuse potential}

\Plots can be used by people working in all fields for data visualization.
In particular it is possible to define backend agnostic recipes for their domain specific data structures with minimal dependencies.
These can be shared, reused and extended by peers with ease by including these recipes in their packages or published scripts.
Also it is possible for other plotting software with julia bindings to take advantage of the recipe system either by contributing backend code to \Plots or by using \inline{RecipesPipeline.jl} to become an independent consumer of \inline{RecipesBase.jl}'s recipes.
Plotting software without julia bindings could potentially use the HDF5 backend to consume fully processed and serialized recipe data.

People interested in modifying, extending or maintaining \Plots can get in contact either via the github issue tracker, the julia discourse forum or the julia slack and zulip spaces.
There are quarterly maintenance calls that can be joined on request.

\section*{Acknowledgements}

We like to acknowledge the support of the julia community and the numerous contributors that keep this project alive.

\section*{Funding statement}

\section*{Competing interests}

The authors have no competing interests to declare.

\section*{Code examples}
\jlinputlisting[caption=Recipes showcase, label = lst:recipe]{./fig/recipe_showcase.jl}

\printbibliography

\end{document}